\documentclass[conference]{IEEEtran}
\usepackage{cite} 
\usepackage{algorithm}
\usepackage{algorithmic}
\usepackage{amssymb}

\ifx\pdfoutput\undefined
\usepackage{graphicx}
\else
\usepackage[pdftex]{graphicx}
\fi

\ifx\pdfoutput\undefined
\usepackage[hypertex]{hyperref}
\else
\usepackage[pdftex,hypertexnames=false]{hyperref}
\fi

\begin{document}

\title{Assessing Random Dynamical Network Architectures for Nanoelectronics}

\author{Christof Teuscher$^1$, Natali Gulbahce$^2$, Thimo Rohlf$^3$\\ $^1$Computer, Computational \& Statistical Sciences Division, Los Alamos National Laboratory, USA, \\ {\tt christof@teuscher.ch}\\
$^2$Center for Complex Networks Research, Northeastern University, USA,\\ {\tt natali.gulbahce@gmail.com} \\ $^3$Max Planck Institute for Mathematics in the Sciences, Leipzig, Germany, \\ {\tt rohlf@santafe.edu}}

\maketitle

\begin{abstract}
  Independent of the technology, it is generally expected that future
  nanoscale devices will be built from vast numbers of densely
  arranged devices that exhibit high failure rates. Other than that,
  there is little consensus on what type of technology and computing
  architecture holds most promises to go far beyond today's top-down
  engineered silicon devices. Cellular automata (CA) have been
  proposed in the past as a possible class of architectures to the von
  Neumann computing architecture, which is not generally well suited
  for future massively parallel and fine-grained nanoscale
  electronics. While the top-down engineered semi-conducting
  technology favors regular and locally interconnected structures,
  future bottom-up self-assembled devices tend to have irregular
  structures because of the current lack of precise control over these
  processes. In this paper, we will assess random dynamical networks,
  namely Random Boolean Networks (RBNs) and Random Threshold Networks
  (RTNs), as alternative computing architectures and models for future
  information processing devices. We will illustrate that---from a
  theoretical perspective---they offer superior properties over
  classical CA-based architectures, such as inherent robustness as the
  system scales up, more efficient information processing
  capabilities, and manufacturing benefits for bottom-up designed
  devices, which motivates this investigation. We will present recent
  results on the dynamic behavior and robustness of such random
  dynamical networks while also including manufacturing issues in the
  assessment.
\end{abstract}

\pagestyle{empty}
\thispagestyle{empty}

\section{Introduction and Motivation}
\label{sec:intro}
The advent of multicore architectures and the slowdown of the
processor's operating frequency increase are signs that CMOS
miniaturization is increasingly hitting fundamental physical limits. A
key question is how computing architectures will evolve as we reach
these fundamental limits. A likely possibility within the realm of
CMOS technology is that the integration density will cease to increase
at some point, instead only the number of components, i.e, the
transistors, will further increase, which will necessarily lead to
chips with a higher area. This trend can already be observed with
multi-core architectures. That in itself has implications on the
interconnect architecture, the power consumption and dissipation, and
the reliability. Another possibility is to go beyond silicon-based
technology and to change the computing and manufacturing paradigms, by
using for example bottom-up self-assembled devices. Self-assembling
nanowires \cite{ferry08} or carbon nanotube electronics
\cite{avouris07} are promising candidates, although none of them has
resulted in electronics that is able to compete with traditional CMOS
so far. What seems clear is that the current way with build computers
and the way we algorithmically solve problems with them may need to be
fundamentally revisited, which this paper is all about.

While the top-down engineered CMOS technology favors regular and
locally interconnected structures, future bottom-up self-assembled
devices tend to have irregular structures because of the current lack
of precise control over these processes. We therefore hypothesize that
future and emerging computing architectures will be much more driven
by manufacturing constraints and particularities than for CMOS, which
allowed engineers to implement a logic-based computing architecture
with extreme precision and reliability, at least in the
past. Independent of the forthcoming device and fabrication
technologies, it is generally expected that future nanoscale devices
will be built from (1) vast numbers of densely arranged devices that
(2) exhibit high failure rates. We take this working hypothesis for
granted in this paper and address it from a perspective that focuses
on the interconnect topology. This is justified by the fact that the
importance of interconnects on electronic chips has outrun the
importance of transistors as a dominant factor of performance
\cite{meindl03,ho01,davis01}. The reasons are twofold: (1) the
transistor switching speed for traditional silicon is much faster than
the average wire delays and (2) the required chip area for
interconnects has dramatically increased.

In \cite{zhirnov08}, Zhirnov et al. explored integrated digital {\sl
  Cellular Automata} (CA) architectures---which are highly regular
structures with local interconnects (see Section \ref{sec:ca})---as an
alternative paradigms to the von Neumann computer architecture for
future and emerging information processing devices. Here, we are
interested to explore and assess a more general class of discrete
dynamical systems, namely {\sl Random Boolean Networks} (RBNs) and
{\sl Random Threshold Networks} (RTNs). We will mainly focus on RBNs,
but RTNs are included in this paper because they offer an alternative
paradigm to Boolean logic, which can be efficiently implemented as
well (see Section \ref{sec:fab}).

Motivated by future and emerging nanoscale devices, we are interested
to provide answers to the following questions:
\begin{itemize}
\item Do RBNs and RTNs offer benefits over CA-architectures? If yes,
  what are they?
\item How does the interconnect complexity compare between RBNs/RTNs
  and CAs?
\item Does any of these architectures allow to solve problems more
  efficiently?
\item Is any of these architectures inherently more robust to simple
  errors?
\item Can CMOS and beyond-CMOS devices provide a benefit for the
  fabrication of any of these architectures?
\end{itemize}

We will argue and illustrate that---at least from a theoretical
perspective---random dynamical networks offer superior properties over
classical regular CA-based architectures, such as inherent robustness
as the system scales up, more efficient information processing
capabilities, and manufacturing benefits for bottom-up fabricated
devices, which motivates this investigation. We will present recent
results on the dynamic behavior and robustness of such random
dynamical networks while also including manufacturing issues in the
assessment. 

To answer the above questions, we will extend recent results on the
complex dynamical behavior of discrete random dynamical networks
\cite{rohlf07_prl}, their ability to solve problems
\cite{mesot05,teuscher07:ddaysrbn}, and novel interconnect paradigms
\cite{teuscher07:chaos,teuscher08:iscas08}.

The remainder of this paper is as following: Section \ref{sec:rdn}
introduces random dynamical networks, namely random Boolean and random
threshold networks. Section \ref{sec:ca} briefly presents cellular
automata architectures. Damage spreading and criticality of cellular
automata and random dynamical networks is analyzed in Section
\ref{sec:damage}. Section \ref{sec:complex} analyzes the network
topologies from a graph-theoretical and wiring-cost perspective. The
task solving capabilities of RBNs and CAs are briefly assessed in
Section \ref{sec:problem}, while Section \ref{sec:fab} looks into
manufacturing issues. Section \ref{sec:conclusion} concludes the
paper.

\section{Random Dynamical Networks}
\label{sec:rdn}

\subsection{Random Boolean Networks}
A {\sl Random Boolean Network} (RBN)
\cite{kauffman69,kauffman84,kauffman93} is a discrete dynamical system
composed of $N$ nodes, also called {\sl automata}, {\sl elements} or
{\sl cells}. Each automaton is a Boolean variable with two possible
states: $\{0,1\}$, and the dynamics is such that

\begin{equation}
{\bf F}:\{0,1\}^N\mapsto \{0,1\}^N, 
\label{globalmap}
\end{equation} 

where ${\bf F}=(f_1,...,f_i,...,f_N)$, and each $f_i$ is represented
by a look-up table of $K_i$ inputs randomly chosen from the set of $N$
nodes. Initially, $K_i$ neighbors and a look-table are assigned to
each node at random.  Note that $K_i$ (i.e., the fan-in) can refer to
the {\sl exact} or to the {\sl average} number of incoming connections
per node.

A node state $ \sigma_i^t \in \{0,1\}$ is updated using its
corresponding Boolean function:

\begin{equation}
\sigma_i^{t+1} = f_i(x_{i_1}^t,x_{i_2}^t, ... ,x_{i_{K_i}}^t).
\label{update}
\end{equation}

These Boolean functions are commonly represented by {\sl
  lookup-tables} (LUTs), which associate a $1$-bit output (the node's
future state) to each possible $K$-bit input configuration. The
table's out-column is called the {\sl rule} of the node. Note that
even though the LUTs of a RBN map well on an FPGA or other
memory-based architectures, the random interconnect in general does
not.

\begin{figure}[htb]
  \centering \includegraphics[width=.47\textwidth]{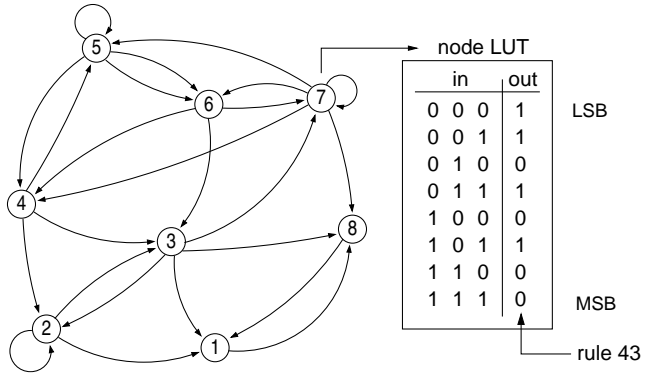}
  \caption{Illustration of a random Boolean network with $N=8$ nodes
    and $K=3$ inputs per node (self-connections are allowed).  The
    node rules are commonly represented by {\sl lookup-tables} (LUTs),
    which associate a $1$-bit output (the node's future state) to each
    possible $K$-bit input configuration. The table's out-column is
    commonly called the {\sl rule} of the node.}
  \label{fig:rbn}
\end{figure}

We randomly initialize the states of the nodes (initial condition of
the RBN). The nodes are updated synchronously using their
corresponding Boolean functions. Other updating schemes exist, see for
example \cite{gershenson2003:alife} for an overview. Synchronous
random Boolean networks as introduced by Kauffman are commonly called
{\sl NK} networks or models. Figure \ref{fig:rbn} shows a possible
{\sl NK} random Boolean network representation ($N=8, K=3$).

\subsection{Random Threshold Networks}
\label{sec:rbn}
{\sl Random Threshold Networks} (RTNs) are another type of discrete
dynamical systems. An RTN consists of $N$ randomly interconnected
binary sites (spins) with states $\sigma_i=\pm1$.  For each site $i$,
its state at time $t+1$ is a function of the inputs it receives from
other spins at time $t$:

\begin{eqnarray} 
\sigma_i(t+1) = \mbox{sgn}\left(f_i(t)\right) 
\end{eqnarray}  
with 
\begin{eqnarray} 
f_i(t) = \sum_{j=1}^N c_{ij}\sigma_j(t) + h.  
\end{eqnarray}

The $N$ network sites are updated synchronously. In the following, the
threshold parameter $h$ is set to zero. The interaction weights
$c_{ij}$ take discrete values $c_{ij} = +1$ or $-1$ with equal
probability. If $i$ does not receive signals from $j$, one has $c_{ij}
= 0$.

\section{Cellular Automata Architectures}
\label{sec:ca}
\begin{figure}[htb]
  \centering \includegraphics[width=.38\textwidth]{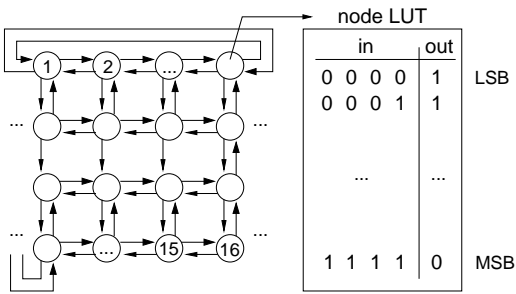}
  \caption{Illustration of a binary, 2D, folded cellular automaton
    with $N=16$ cells. Each node is connected to its four immediate
    neighbors (von Neumann neighborhood).}
  \label{fig:2dca}
\end{figure}

{\sl Cellular automata} (CA) \cite{wolfram84} were originally
conceived by Ulam and von Neumann \cite{neumann66:_theor} in the 1940s
to provide a formal framework for investigating the behavior of
complex, extended systems. CAs are a special case of the more general
class of random dynamical networks, in which space and time are
discrete. A CA usually consists of a $D$-dimensional regular lattice
of $N$ lattice sites, commonly called {\sl nodes}, {\sl cells}, {\sl
  elements}, or {\sl automata}. Each cell $i$ can be in one of a
finite number of $S$ possible states and further consists of a
transition function $f_i$ (also called {\sl rule}), which maps the
neighboring states to the set of cell states. CAs are called {\sl
  uniform} if all cells contain the same rule, otherwise they are {\sl
  non-uniform}.  Each cell takes as input the states of the cells
within some finite local neighborhood. Here, we only consider
non-uniform, two-dimensional ($D=2$), folded, and binary CAs ($S=2$)
with a radius-$1$ von Neumann neighborhood, where each cell is
connected to each of its four immediate neighbors only. Figure
\ref{fig:2dca} illustrates such an CA. The Boolean functions in each
node must therefore define $2^4=16$ possible input combinations. To be
able to compare CAs with RBNs, we do not consider self-connections.

\section{Damage Spreading and Criticality}
\label{sec:damage}

\subsection{Random Boolean and Threshold Networks}
As we have seen in Section \ref{sec:rbn}, RBNs and their complex
dynamic behavior are essentially characterized by the average number
of incoming links $K_i$ (fan-in) per node (e.g., Figure \ref{fig:rbn}
shows a $K=3$ network with 3 incoming links per node). It turns out
that in the thermodynamic limit, i.e., $N \rightarrow \infty$, RBNs
exhibit a dynamical order-disorder transition at a sparse critical
connectivity $K_c=2$ \cite{derrida86} (i.e., where each node receives
on average two incoming connections from two randomly chosen other
nodes), which partitions their operating space into 3 different
regimes: (1), sub-critical, where $\langle K \rangle < K_c$, (2)
complex, where $\langle K \rangle = K_c$, and (3) supercritical, where
$\langle K \rangle > K_c$. In the sub-critical regime, the network
dynamics are too ``rigid'' and the information processing capabilities
are thus hindered, whereas in the supercritical regime, their behavior
becomes chaotic. The complex regime is also commonly called the ``edge
of chaos,'' because it represents the network connectivity where
information processing is ``optimal'' and where a small number of
stable attractors exist.

Similar observations were made for sparsely connected random threshold
(neural) networks (RTN) \cite{rohlf02} for $K_c=1.849$.  For a finite
system size $N$, the dynamics of both systems converge to periodic
attractors after a finite number of updates. At $K_c$, the phase space
structure in terms of attractor periods \cite{albert00_prl}, the
number of different attractors \cite{samuelsson03} and the
distribution of basins of attraction \cite{bastolla98} is complex,
showing many properties reminiscent of biological networks
\cite{kauffman93}.

\paragraph*{Results} 
In \cite{rohlf07_prl} we have systematically studied and compared
damage spreading (i.e., how a perturbed node-state influences the rest
of the network nodes over time) at the sparse percolation (SP) limit for
random Boolean and threshold networks with perturbations. In the SP
limit, the damage induced in a network (i.e., by changing the state of
a node) does not scale with system size. Obviously, this limit is
relevant to information and damage propagation in many technological
and natural networks, such as the Internet, disease spreading in
populations, failure propagation in power grids, and
networks-on-chips. We measure the damage spreading by the following
methodology: the state of one randomly chosen node is changed. The
damage is measured as the Hamming distance between a damaged and
undamaged network instance after a large number of $T$ system updates.

\begin{figure}
  \centering \includegraphics[width=0.47\textwidth]{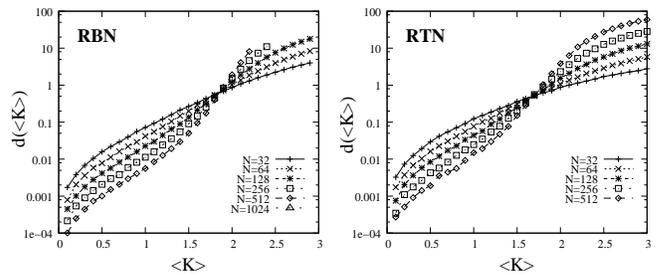}
  \caption{Average Hamming distance (damage) $\langle d \rangle$ after
    200 system updates, averaged over 10000 randomly generated
    networks for each value of $\langle K \rangle$, with 100 different
    random initial conditions and one-bit perturbed neighbor
    configurations for each network. For both RBN and RTN, all curves
    for different $N$ approximately intersect in a characteristic
    point $K_s$.}
  \label{fig:dofk_rbnrtn_f}
\end{figure}

We have shown that there is a characteristic average connectivity
$K_s^{RBN}=1.875$ for RBNs and $K_s^{RTN}=1.729$ for RTNs, where the
damage spreading of a single one-bit perturbation of a network node
remains constant as the system size $N$ scales up. Figure
\ref{fig:dofk_rbnrtn_f} illustrates this newly discovered point for
RBNs and RTNs. For more details, see \cite{rohlf07_prl}.

\paragraph*{Discussion} 
Both $K_c$ and $K_s$ are highly relevant for nano-scale electronics
for the following reason: assuming we can build massive numbers of $N$
simple logic gates that implement a random Boolean function, the above
findings tell us that on average, every gate should be connected
somewhere close to both $K_s$ and $K_c$ in order to (1) guarantee
optimal robustness against failures for any system size and (2)
optimal information processing at the ``edge of chaos.'' We are also
hypothesizing that natural systems, such as the brain or genetic
regulatory networks, may have evolved towards these characteristic
connectivities. This remains, however, to be proved and is part of
ongoing research.

\subsection{Cellular Automata Damage Spreading}
\label{sec:cadamage}
We have used the same approach as described above to measure the
damage spreading in cellular automata. In order to vary the average
number $\langle K \rangle$ of incoming links per cell in a cellular
automata (e.g., as pictured in Figure \ref{fig:2dca}), we have adopted
the following methodology: (1) for a desired average number of links
per cell $\langle K \rangle$ for a given CA size of $N$ cells, the
total number of links in the automaton is given by $L = N\langle K
\rangle$; (2) we then randomly choose $L$ possible connections on the
regular CA-grid with uniform probability and establish the
links. Damage is induced in the same way as for RBNs and RTNs: the
state of one (or several) randomly chosen node(s) is changed. The
damage is measured as the Hamming distance between a damaged and
undamaged CA instance after a large number of $T$ system updates, in
our case $T=200$.

\begin{figure}
  \centering \includegraphics[width=0.47\textwidth]{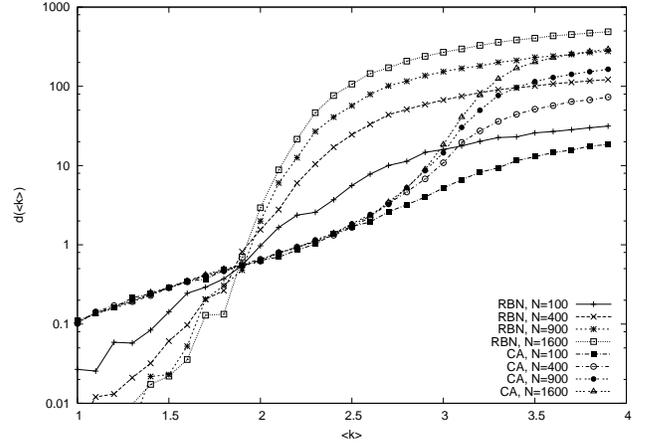}
  \caption{Average Hamming distance (damage) $\langle d \rangle$ after
    200 system updates, averaged over 100 randomly generated networks
    for each value of $\langle K \rangle$, with 100 different random
    initial conditions and a damage size of 1 node for each
    network. See text for discussion.}
  \label{fig:damage_average_k_d1}
\end{figure}

\begin{figure}
  \centering \includegraphics[width=0.47\textwidth]{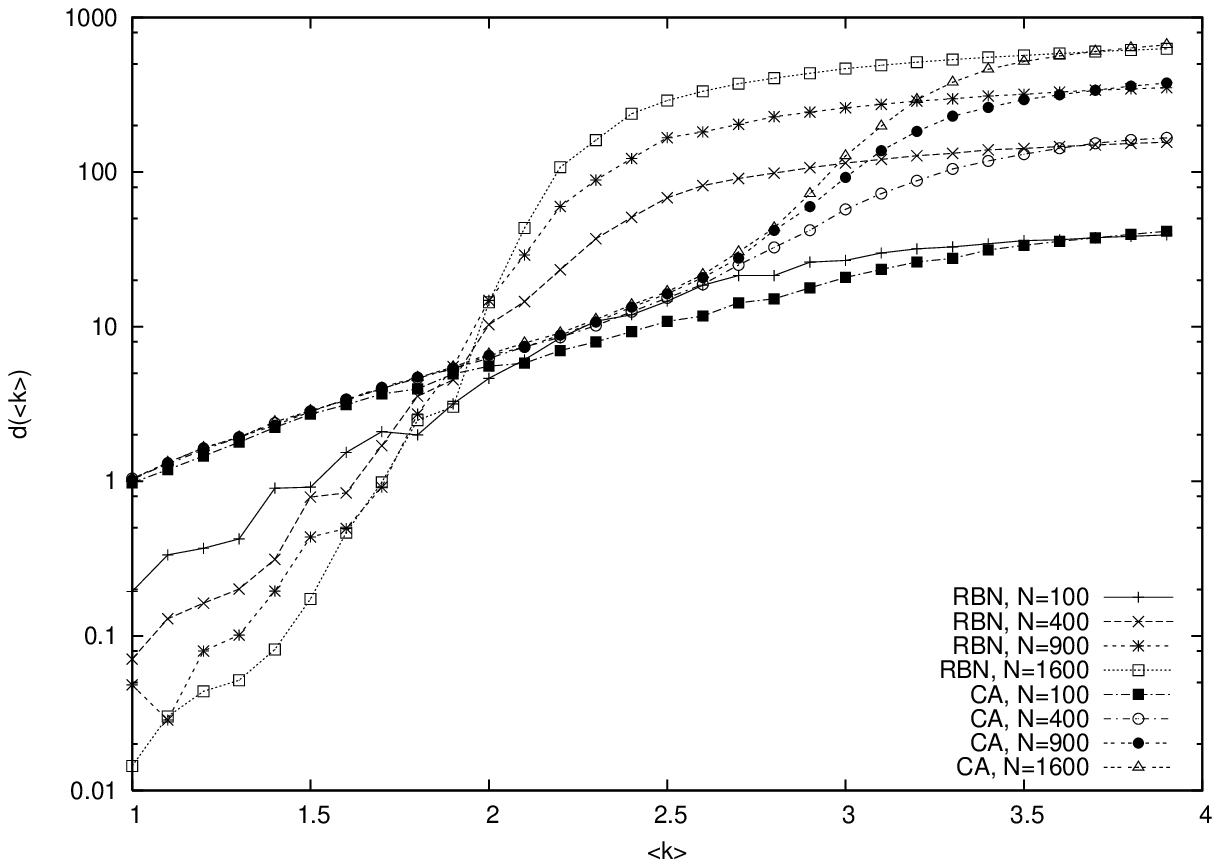}
  \caption{Average Hamming distance (damage) $\langle d \rangle$ after
    200 system updates, averaged over 100 randomly generated networks
    for each value of $\langle K \rangle$, with 100 different random
    initial conditions and a damage size of 10 nodes for each
    network. See text for discussion.}
  \label{fig:damage_average_k_d10}
\end{figure}

\begin{figure}
  \centering \includegraphics[width=0.47\textwidth]{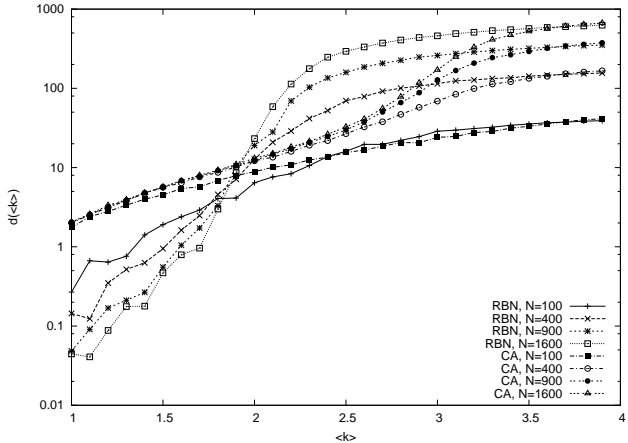}
  \caption{Average Hamming distance (damage) $\langle d \rangle$ after
    200 system updates, averaged over 100 randomly generated networks
    for each value of $\langle K \rangle$, with 100 different random
    initial conditions and a damage size of 20 nodes for each
    network. See text for discussion.}
  \label{fig:damage_average_k_d20}
\end{figure}

\paragraph*{Results}
Figures \ref{fig:damage_average_k_d1}, \ref{fig:damage_average_k_d10},
and \ref{fig:damage_average_k_d20} show the average damage of both
RBNs and CAs for different system sizes and for a damage size of 1 and
10 respectively. We have left out RTNs for this analysis. As one can
see, both the RBN and the CA average damage for different $N$
approximately intersect in the characteristic point
$K_s^{RBN}=1.875$. This point is less pronounced for the larger damage
sizes (Figures \ref{fig:damage_average_k_d10} and
\ref{fig:damage_average_k_d20}). The RBN curves confirm what was
already shown above in Figure \ref{fig:dofk_rbnrtn_f}, and are merely
plotted here for comparison with the CA architectures and their system
sizes imposed by square lattices.

Interestingly, the CAs show different damage propagation behavior for
different system sizes and connectivities. First, we observe that the
average damage for one-bit damage events (Figure
\ref{fig:damage_average_k_d1}) is independent of the system size $N$
for up to approximatively $\langle K \rangle=2.5$ average incoming
connections per cell. This behavior disappears completely for large
damage sizes (Figure \ref{fig:damage_average_k_d20}). Second, Figure
\ref{fig:damage_average_k_d1} shows that all curves intersect at
$K_s^{RBN}=K_s^{CA}=1.875$. Third, Figure
\ref{fig:damage_average_k_d20} suggest that for larger damage sizes,
$K_s^{CA}$ disappears for CAs. Fourth, the average damage for larger
damage events, i.e., 10 and 20 in our examples, converges to the same
final values for both RBNs and CAs as $\langle K \rangle$ approaches
4.

\paragraph*{Discussion} 
We hypothesize that the particular behavior can be explained by the
percolation limit of the cellular automata. Da Silva et
al. \cite{dasilva89} found that the link probability at the
percolation limit is approximatively $p \sim 0.6$, which means that
the average connectivity at the percolation limit in our CA topology
with a maximum of 4 neighbors is given by $\langle k \rangle = 4p =
2.4$. This value corresponds to the experimentally observed value
where the damage spreading suddenly becomes dependent of the system
size.  Because of the local CA connectivity, there are lots of
disconnected components below the percolation limit. Below this limit,
the damage spreading is thus very slow and limited by the disconnected
components, reason why it is essentially independent of system
size. Above the percolation limit, the CA suddenly becomes connected
and damage spreading becomes therefore dependent on the system size.
For larger damage events, such as 10 or 20, damage becomes more
dependent on system size even below the percolation limit because
there is a higher probability that damage is induced in several
disconnected components at the same time.

In summary: for single-node damage events, CAs offer system-size
independent damage spreading for up to about $\langle K \rangle=2.4$
(which corresponds to the percolation limit), however, this particular
behavior disappears for larger damage events. We conclude that in the
general case, CAs do not possess a characteristic connectivity $K_s$,
where damage spreading is independent of the system size $N$. Such a
connectivity, however, exists for both RBNs and RTNs, which makes them
particularly suitable as a computing model in an environment with high
error probabilities or systems with low system component
reliabilities. An example are logical gates based on bio-molecular
components \cite{benenson2001}, where high failure rates can be
expected.

\section{Complex Networks and Wiring Costs}
\label{sec:complex}
Most real networks, such as brain networks \cite{sporns04,egueluz05},
electronic circuits \cite{cancho01}, the Internet, and social networks
share the so-called {\em small-world} (SW) property
\cite{watts98}. Compared to purely locally and regularly
interconnected networks (such as for example the CA interconnect of
Figure \ref{fig:2dca}), small-world networks have a very short average
distance (measured as the number of edges to traverse) between any
pair of nodes, which makes them particularly interesting for efficient
communication.

The classical Watts-Strogatz small-world network \cite{watts98} is
built from a regular lattice with only nearest neighbor
connections. Every link is then rewired with a {\em rewiring
  probability} $p$ to a randomly chosen node. Thus, by varying $p$,
one can obtain a fully regular ($p=0$) and a fully random ($p=1$)
network topology. The rewiring procedure establishes ``shortcuts'' in
the network, which significantly lower the average distance (i.e., the
number of edges to traverse) between any pair of nodes. In the
original model, the length distribution of the shortcuts is uniform
since a node is chosen randomly. If the rewiring of the connections is
done proportional to a power law, $l^{-\alpha}$, where $l$ is the wire
length, then we obtain a {\em small-world power-law network}. The
exponent $\alpha$ affects the network's communication characteristics
\cite{kozma05} and navigability \cite{kleinberg00}, which is better
than in the uniformly generated small-world network. One can think of
other distance-proportional distributions for the rewiring, such as
for example a Gaussian distribution, which has been found between
certain layers of the rat's neocortical pyramidal neurons
\cite{hellwig00}. 

In a real network, it is fair to assume that local connections have a
lower cost (in terms of the associated wire-delay and the area
required) than long-distance connections.  Physically realizing
small-world networks with uniformly distributed long-distance
connections is thus not realistic and distance, i.e., the wiring cost,
needs to be taken into account, a perspective that recently gained
increasing attention \cite{petermann06}. On the other hand, a
network's topology also directly affects how efficient problems can be
solved.

Teuscher \cite{teuscher07:chaos} has pragmatically and experimentally
investigated important design trade-offs and properties of an
irregular, abstract, yet physically plausible 3D small-world
interconnect fabric that is inspired by modern network-on-chip
paradigms. The results confirm that (1) computation in irregular
assemblies is a promising and disruptive computing paradigm for
self-assembled nano-scale electronics and (2) that 3D small-world
interconnect fabrics with a power-law decaying distribution of
shortcut lengths are physically plausible and have major advantages
over local 2D and 3D regular topologies, such as CA interconnects.

\paragraph*{Discussion}
There is a trade-off between (1) the physical realizability and (2)
the communication characteristics for a network topology. A locally
and regularly interconnected topology, such as that of a CA, is in
general easy to build (especially for to-down engineered CMOS
technology) and only involves minimal wire and area cost (as for
example shown by Zhirnov et al. \cite{zhirnov08}), but it offers poor
global communication characteristics and scales-up poorly with system
size. On the other hand, a random topology, such as that of RBNs or
RTNs, scales-up well and has a very short-average path length, but it
is not physically plausible because it involves costly long-distance
connections established independently of the Euclidean distance
between the nodes. The RBN and RTN topologies we consider here as thus
extremes, such as CA topologies, the ideal lies in between:
small-world topologies with a distance-dependent distribution of the
connectivity. Such topologies are located in a unique spot in the
design space and also offer two other highly relevant properties
\cite{kleinberg00_acm,teuscher07:chaos}: (1) efficient navigability
and thus potentially efficient routing, and (2) robustness against
random link removals. For these reasons, we can conclude that
small-world graphs are the most promising interconnects for future
massive scale devices.

\section{A Glance on Task Solving}
\label{sec:problem}
In \cite{mesot05}, Mesot and Teuscher have presented a novel
analytical approach to find the local rules of random Boolean networks
to solve the global density classification and the synchronization
task---which are well known benchmark tasks in the CA community---from
any initial configuration. They have also quantitatively and
qualitatively compared the results with previously published work on
cellular automata and have shown that randomly interconnected automata
are computationally more efficient in solving these two global tasks.

In addition, preliminary results by the authors
\cite{teuscher07:ddaysrbn} also suggest that $K_c=2$ RBN generalize
better on simple learning tasks than sub-critical or supercritical
networks, but more research will be necessary.

\paragraph*{Discussion}
To efficiently solve algorithmic problems with distributed computing
architectures, efficient communication is key. This is particularly
true for tasks such as the density or the synchronization task, which
are trivial to solve if one has a global view on the entire system
state, but non-trivial to solve if each cell only sees a limited
number of neighboring cells. It is thus not surprising that cells
interconnected by a network with the small-world property perform much
better on such tasks because the information propagation is
significantly better. This is a too often neglected fact for CAs, in
particular if one wants to use them as a viable mainstream and general
purpose computing architecture.  It is well-know that even simple CAs
are computationally universal (and so are RBNs), i.e., they can solve
any algorithmic problem, but due to their local non-small-world
interconnect topology, that will only be possible in a highly
inefficient way in the general case, i.e., for a large set of
different applications. This is well illustrated with the (highly
inefficient) implementation of a universal Turing machine on top of
the {\em Game of Life} \cite{rendell02}. Naturally, there are
exceptions to the general case, and it has been shown that CAs can be
extremely efficient for certain niche applications, such as for examle
image processing.

\section{Manufacturing Issues}
\label{sec:fab}
As Chen et al. \cite{chen03} state, ``[i]n order to realize functional
nano-electronic circuits, researchers need to solve three problems:
invent a nanoscale device that switches an electric current on and
off; build a nanoscale circuit that controllably links very large
numbers of these devices with each other and with external systems in
order to perform memory and/or logic functions; and design an
architecture that allows the circuits to communicate with other
systems and operate independently on their lower-level details.''

While we can currently build switching devices in various technologies
besides CMOS (see \cite{zhirnov06,bourianoff08,hutchby08} for an
overview), one of the remaining challenges is to assemble and
interconnect these switching devices (or logic functions) to larger
systems, and ultimately to design a computing architecture that allows
to perform reliable computations. As mentioned before, there is little
consensus in the research community on what type of technology and
computing architecture holds most promises for the future.

The motivation for investigating randomly assembled interconnects and
computing architectures can be summarized by the following
observations:

\begin{itemize}
  \item long-range and global connections are costly (in terms of wire
    delay and of the chip area used) and limit system performance
    \cite{ho01};
  \item it is unclear whether a precisely regular and homogeneous
    arrangement of components is needed and possible on a
    multi-billion-component or even Avogadro-scale assembly of
    nano-scale components \cite{tour02}
  \item ``[s]elf-assembly makes it relatively easy to form a random
    array of wires with randomly attached switches'' \cite{zhirnov01};
    and
  \item building a perfect system is very hard and expensive
\end{itemize}

We have hypothesized in \cite{teuscher08:iscas08} and
\cite{teuscher07:chaos} that bottom-up self-assembled electronics
based on conductive nanowires or nanotubes can lead to the random
interconnect topologies we are interested in, however, several
questions remain open and are part of a 3-year interdisciplinary
research project at LANL. Our approach consists in using a hybrid
assembly (as others explore as well, e.g., \cite{ferry08}), where the
functional building blocks will still be traditional silicon in a
first step, while the interconnect is made up from self-assembled
nanowires. Nanowires can be grown in various ways using diverse
materials, such as metals and semiconductors. We have chosen a novel
way to grow conductive nanowires, which Wang et al. \cite{wang07} at
LANL have pioneered and demonstrated: Ag nanowires can be fabricated
on top of conducting polyaniline polymer membranes via a spontaneous
electrodeless deposition (self-assembly) method. We hypothesize that
this will allow to densely interconnect silicon components in a simple
and cheap way with specific distance-dependent wire-length
distributions. We believe that this approach will ultimately allow us
to easily and cheaply fabricate RBN-like computing architectures.

Random threshold networks, on the other hand, could be rather
straightforwardly and efficiently implemented with resonant tunneling
diode (RTD) logic circuits (see e.g., \cite{pettenghi08}), and
represent a very interesting alternative to conventional Boolean logic
gates.  The reported results in this paper on random threshold
networks can thus directly be applied to the implementation of such
devices. There has been a significant body of research in the area of
threshold logic in the past (see e.g., \cite{muroga71}), but to the
best of our knowledge, random threshold networks have not been
considered as computing models for future and emerging computing
machines.

\section{Conclusion}
\label{sec:conclusion}
The central claim of this paper is that locally interconnected
computing architectures, such as cellular automata (CA), are in
general not appropriate models for large-scale and general-purpose
computations. We have supported this claim with recent theoretical
results on the complex dynamical behavior of discrete random dynamical
networks, their robustness to damage events as the system scales up,
their ability to efficiently solve tasks, and their improved transport
characteristics due to the short average path length. The arguments,
in a nutshell, why we believe that CAs are {\sl not} promising
architectures for future information-processing devices, are as
following:

\begin{itemize}
\item their local interconnect topology is not small-world and has
  thus worse global transport characteristics (than small-world or
  random graphs), which directly affects the effectiveness of how
  general-purpose algorithmic tasks can be solved;
\item in terms of a complex dynamical system, they operate in the
  supercritical regime ($\langle K \rangle > K_c$) with the widely
  used von Neumann neighborhood, which makes them sensitive to initial
  conditions;
\item they do not generally have a characteristic connectivity $K_s$,
  where damage spreading is independent of system size, which makes a
  system inherently robust; and
\item it is unclear whether a precisely regular and homogeneous
  arrangement of components is possible at the scale of future
  information processing devices.
\end{itemize}

We have assessed RBNs and RTNs as alternative models, however, as we
have seen in Section \ref{sec:complex} they come at a serious cost:
the uniform probability to establish connections with any node in the
system independent of the Euclidean distance between them is not
physically plausible and too expensive in terms of wiring cost. The
ultimate interconnect topology is small-world and has a
distance-dependent distribution of the wires
\cite{teuscher07:chaos,teuscher08:iscas08,petermann06}. We have
preliminary evidence that, if we were to connect RBNs and RTNs by such
a network topology, both $K_c$ and $K_s$ would still exist. Research
to clarify this question is under progress,

\paragraph*{Open Questions and Unaddressed Issues}
Naturally, there are a number of open questions and issues that we
have not addressed because they are beyond the scope of this paper. In
particular, an irregular topology with random logical functions makes
the mapping of a given digital circuit much harder, if not impossible
in certain cases. On the other hand, a regular interconnect topology
clearly makes the mapping task easier. We believe, however, that this
challenge can be addressed by automated design tools. After all,
computation in random assemblies is not completely new and has been
more or less successfully tried by others, e.g.,
\cite{patwardhan06,lawson06,tour02}, however in different contexts and
with a different perspective in mind than we have presented here.

We have deliberately {\em not} focused on any particular application
in this paper because our results are independent of the
application. However, it is noteworthy that locally interconnected CAs
have been proven to outperform other general purpose architecture on
very specific applications. A good example are cellular neural
networks (CNNs) \cite{chua02}, which, e.g., allow to perform certain
imagine processing tasks orders of magnitude faster than any other
machine.

Further, it is unknown at this point how exactly our findings fit into
the interconnect predictions made by Rent's rule, however, the rule
may not be applicable to our non-traditional circuits since it is
based on empirical results. Further research on this is planned.

Last but not least, we would like to mention that, although we have
only considered 2D arrangements and interconnects here for simplicity,
the future is clearly 3D (e.g., see \cite{pavlidis07}). The main
reason is that the average wire length in 3D is shorter than in 2D
interconnects.

\paragraph*{Outlook}
We believe that computation in random self-assemblies of simple
components and interconnections is a highly appealing paradigm, both
from the perspective of fabrication as well as performance and
robustness. Future work will focus on (1) the manufacturing issues,
(2) appropriate design methodologies, (3) addressing the mapping
issues, and (4) more realistic models, which will allow to better
assess the performance and cost, and (5) specific applications.

\subsection*{Acknowledgments}
We gratefully acknowledge the support of the U.S. Department of Energy
through the LANL/LDRD Program for this work. The authors would like to
thank Elshan A. Akhadov and Hsing-Lin Wang.

\end{document}